\documentclass[12pt,a4paper]{article}
\pdfoutput=1
\usepackage{lmodern}
\usepackage{jheppub}	
\usepackage{graphicx}				
\usepackage{amsmath}
\usepackage{mathtools}
\usepackage{amssymb}
\usepackage{booktabs}
\usepackage{comment}

\usepackage{overpic}
\usepackage[caption=false]{subfig}
\usepackage{array}
\usepackage{rotating}

\usepackage[T1]{fontenc} 

\usepackage{color}
\definecolor{dark-gray}{gray}{0.20}
\definecolor{gray}{gray}{0.30}
\definecolor{light-gray}{gray}{0.80}
\definecolor{dark-red}{rgb}{0.7,0,0}
\definecolor{dark-green}{rgb}{0.1,0.4,0}
\definecolor{dark-blue}{rgb}{0.3,0.3,0.7}
\definecolor{light-blue}{rgb}{0.8,0.8,1}

\usepackage{hyperref}
\hypersetup{
	colorlinks=true,
	linkcolor=dark-blue,
	citecolor=dark-red,
	urlcolor=dark-green,
	linktoc=page
}

\newcommand{\dd}{\mathrm{d}}

\newcommand{\e}{\mathrm{e}}

\newcommand{\be}{\begin{equation}}
\newcommand{\ee}{\end{equation}}
\newcommand{\bea}{\begin{eqnarray}}
\newcommand{\eea}{\end{eqnarray}}

\newcommand{\f}[2]{\frac{#1}{#2}}

\title{Page Curve for an Evaporating Black Hole}
					\author{\footnotesize{F}ri{\dh}rik Freyr Gautason$^{\diamond\star}$,}
                                           \author{Lukas Schneiderbauer$^{\star}$,}
                                           \author{Watse Sybesma$^{\star}$}
                                           \author{and L\'{a}rus Thorlacius$^{\star}$}
                                           \affiliation{$^{\diamond}$Instituut voor Theoretische Fysica, KU Leuven\\
Celestijnenlaan 200D, 3001 Leuven, Belgium}                                     
                                           \affiliation{$^{\star}$Science Institute
                                           University of Iceland \\Dunhaga 3, 107 Reykjav\'{i}k, Iceland.}
                                           \emailAdd{ffg@kuleuven.be}
                                           \emailAdd{lukas.schneiderbauer@gmail.com}
                                           \emailAdd{watse@hi.is}
                                           \emailAdd{lth@hi.is}
\abstract{
A Page curve for an evaporating black hole in asymptotically flat spacetime is computed by adapting the Quantum Ryu-Takayanagi (QRT) proposal to an analytically solvable semi-classical two-dimensional dilaton gravity theory. The Page time is found to be one third of the black hole lifetime, at leading order in semi-classical corrections. A Page curve is also obtained for a semi-classical eternal black hole, where energy loss due to Hawking evaporation is balanced by an incoming energy flux.
}

\begin{document}
\maketitle

\section{Introduction}
If black hole evaporation is a unitary process, the entanglement entropy between the outgoing radiation and the quantum state associated to the remaining black hole is expected to follow the so-called Page curve as a function of time~\cite{Page:1993wv,Page:2013dx}. Early on, the entanglement entropy is then a monotonically increasing function of time which closely tracks the coarse grained thermal entropy of the radiation that has been emitted up to that point. This changes when the coarse grained entropy of the radiation exceeds the coarse grained entropy associated to the remaining black hole, at which point the entanglement entropy is limited by the black hole entropy and becomes a decreasing function of time. The time when the entanglement entropy transitions from increasing to decreasing is referred to as the Page time. Reproducing the Page curve without explicitly assuming unitarity is an important step towards resolving Hawking's black hole information paradox~\cite{Hawking:1976aa}.

In a recent breakthrough, a Page curve was computed using semi-classical methods  by studying black holes in asymptotically anti-de Sitter (AdS) spacetimes coupled to a conformal field theory (CFT) reservoir~\cite{Penington:2019npb,Almheiri:2019psf}.\footnote{See {\it e.g.} \cite{Rocha:2008fe} for further details on such a setup.} The result hinges on the use of the Quantum Ryu-Takayanagi (QRT) formula \cite{Ryu:2006bv,Hubeny:2007xt,Faulkner:2013ana,Engelhardt:2014gca} and the existence of extremal hypersurfaces terminating on so-called islands behind the event horizon~\cite{Almheiri:2019hni}. A version of the Page curve can also be obtained for eternal AdS black holes, but in this case the islands extend outside the horizon~\cite{Almheiri:2019yqk}. Explicit computations have for the most part been restricted to two-dimensional Jackiw-Teitelboim gravity~\cite{Almheiri:2019psf,Almheiri:2019hni}, but see \cite{Almheiri:2019psy} for a discussion of islands in higher dimensional AdS black hole spacetimes. 

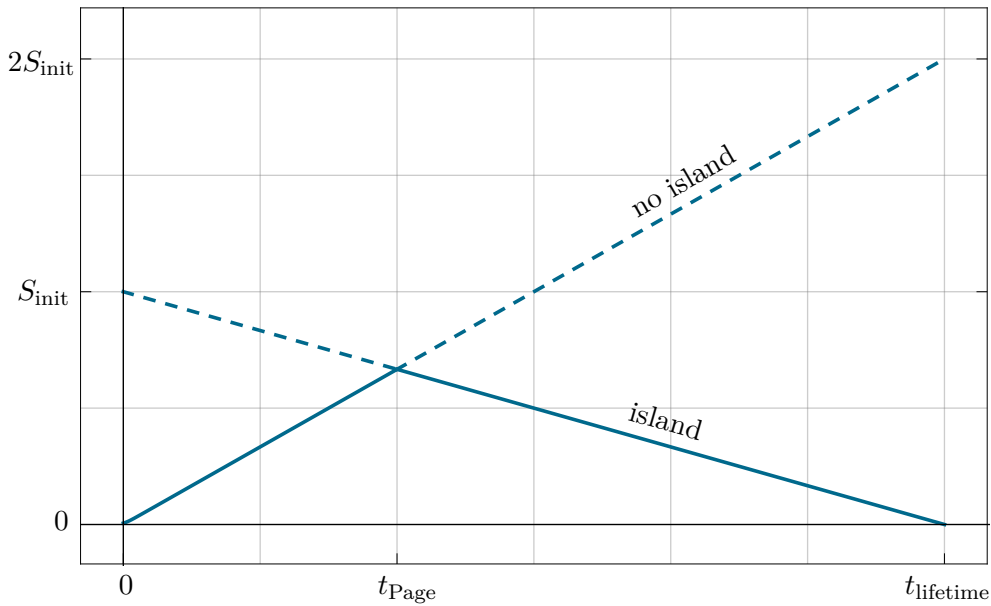
\begin{figure}[t]
    \centering
    \vspace{0.5cm}
    \begin{overpic}[]{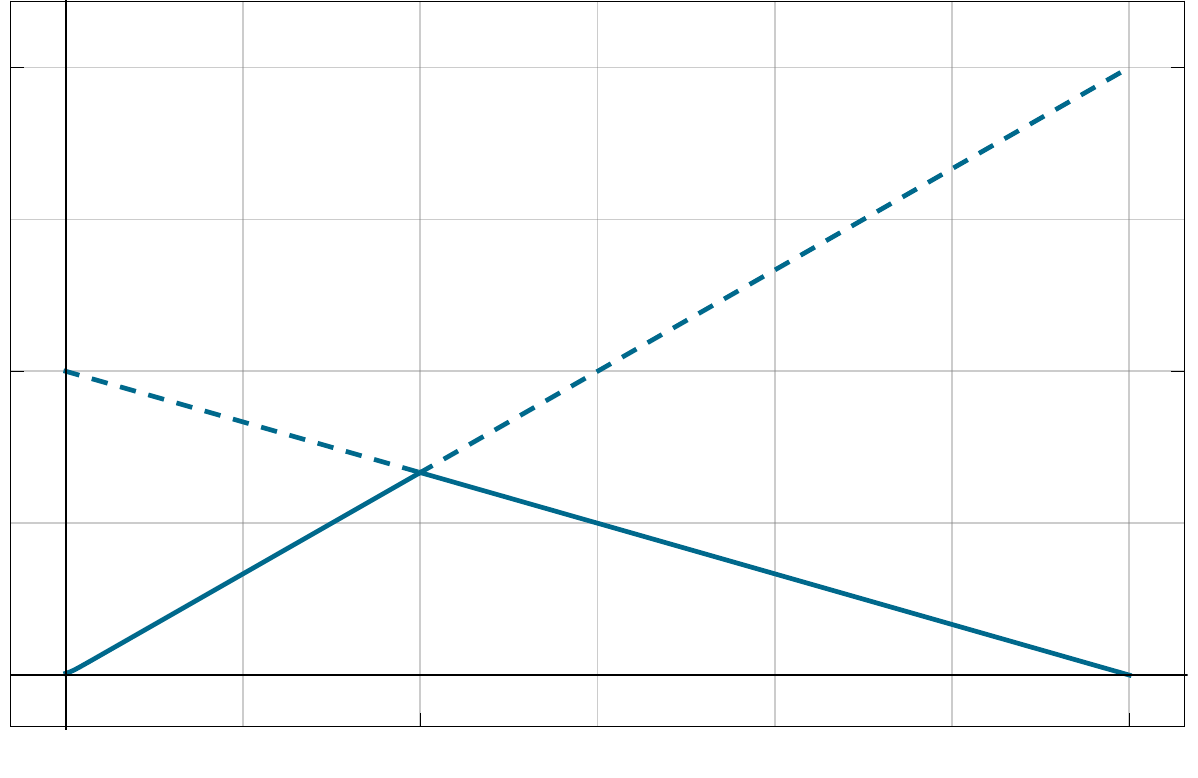}
        \put (5,1) {\small{$0$}}
        \put (33,1) {\small{$t_\text{Page}$}}
        \put (90,1) {\small{$t_\text{lifetime}$}}
        \put (-2,8) {\small{$0$}}
        \put (-6,33) {\small{$S_\text{init}$}}
        \put (-7,58) {\small{$2 S_\text{init}$}}
        \put (60,42) {\rotatebox{30}{\small{no island}}}
        \put (60,20) {\rotatebox{-16}{\small{island}}}
    \end{overpic}
    \caption{\label{backreactpage} Page curve for an evaporating RST black hole.}
    \vspace{0.5cm}
\end{figure}

In the present paper we demonstrate that the QRT formula can also be applied in the context of an evaporating black hole in asymptotically flat spacetime. At leading semi-classical order in the model that we use, and for a large initial black hole mass, the Page time is found to be one third of the black hole lifetime. This main result is presented in Figure~\ref{backreactpage}, where the entanglement entropy of the outgoing Hawking radiation that has passed beyond a distant spatial reference point is plotted as a function of time registered at the reference point. 

We work with a two-dimensional dilaton gravity model of the type introduced by Callan, Giddings, Harvey, and Strominger (CGHS) in \cite{Callan:1992rs}. More specifically, the dilaton gravity sector is that of the model introduced by Russo, Susskind, and Thorlacius (RST) in \cite{Russo:1992ax}, which remains analytically solvable at the semi-classical level. For the matter sector, we take a two-dimensional CFT with a large central charge $c\gg 1$, but rather than working with a large number of free scalar fields as in the CGHS-model, we assume that the conformal matter is holographic. This allows us to take advantage of an insight put forward by Almheiri {\it et al.} \cite{Almheiri:2019hni} in the context of two-dimensional AdS gravity, and use a three-dimensional gravitational dual description to calculate the contribution of the two-dimensional bulk matter to the generalized entropy in the QRT formula. 
In this model, the formation of a black hole from collapsing CFT matter into vacuum and its subsequent evaporation can be studied analytically. The Hawking radiation emitted in this process naturally propagates towards future null infinity. This is in contrast to the AdS setup, where the coupling to a heat bath is essential for letting the radiation escape. Our computation in asymptotically flat spacetime thus gives rise to a rather clean physical picture, both from a computational and conceptual point of view, and offers evidence that the QRT prescription applies beyond asymptotically AdS spacetimes.

The QRT prescription can be motivated via a replica trick involving wormholes, as explored in \cite{Almheiri:2019qdq,Penington:2019kki,Akers:2019nfi,Balasubramanian:2020hfs,Bhattacharya:2020ymw}. Furthermore, the Page curve has been computed from boundary conformal field theories in \cite{Rozali:2019day} and the effect of a quench protocol on the Page curve was studied in \cite{Chen:2019uhq}. Outside the direct scope of black hole physics, Page curves are found to be connected to the eigenstate thermalization hypothesis \cite{Pollack:2020gfa} and the study of chaos \cite{Liu:2020gnp}. 

The paper is organised as follows. In Section \ref{sec:qrt} we review the main result of \cite{Penington:2019npb,Rocha:2008fe,Almheiri:2019hni} and argue for the validity of the QRT formula in our asymptotically flat spacetime model. Then in Section \ref{model} we introduce the two-dimensional dilaton gravity model and establish notation. Following that, in Section \ref{Sgen} we derive the QRT formula for our model. Results for eternal black holes are presented in Section \ref{sec:eternal} and for a dynamical black hole in Section \ref{sec:dynamical}. In Section \ref{sec:conclusion} we present some conclusions and outlook.

\section{Page curve from QRT}\label{sec:qrt}
Our goal is to compute Page curves for black holes in asymptotically flat spacetime.   We will do the computation both for an eternal black hole in an asymptotically linear dilaton spacetime and for a dynamical black hole formed by the gravitational collapse of matter into a linear dilaton vacuum. 
We follow the holographic approach of \cite{Penington:2019npb,Almheiri:2019psf}, which uses the quantum Ryu-Takayanagi (QRT) formula
\be\label{Sgendef}
S_\text{gen} = \f{\text{Area}(I)}{4G_N} + S_\text{Bulk}[{\cal S}_{AI}] \,.
\ee
The first term on the right hand side is the standard Ryu-Takayanagi entropy associated to a given subregion $A$ of the spatial manifold on which the CFT in question is defined. Here $I$ denotes a codimension two region that penetrates into the dual bulk spacetime and is homologous to $A$. The second term is the von Neumann entropy of bulk quantum fields with support inside a spacelike region bounded by $A\cup I$.

In \cite{Penington:2019npb} the system consists of a standard holographic CFT, with Hilbert space ${\cal H}_\text{CFT}$, at finite temperature $T$ so that the dual geometry is an asymptotically AdS black hole. The CFT is assumed to be coupled to an auxiliary system, denoted by ${\cal H}_\text{rad}$, where Hawking radiation emanating from the black hole is collected. The region $A$ in \eqref{Sgendef} above is taken to be the entire boundary where the CFT is defined, {\it i.e.} ${\cal H}_A={\cal H}_\text{CFT}$ and ${\cal H}_{\bar A} = {\cal H}_\text{rad}$, and one looks for regions $I$, homologous to $A$, for which the generalized entropy \eqref{Sgendef} takes extremal values. The QRT prescription for entanglement entropy between $A$ and ${\bar A}$, or in this case the entanglement entropy between the black hole and the Hawking radiation, is then given by the smallest extremal value of \eqref{Sgendef}. At early times the minimum value corresponds to the ``empty'' surface $I=\varnothing$ and the generalized entropy $S_\text{gen}$ is dominated by the von Neumann entropy of the Hawking radiation which grows monotonically with time. Eventually another extremum, involving  an ``island'' $I$ which lies just inside the horizon, takes over.\footnote{More precisely, the island refers to the internal entanglement wedge defined by $I$ \cite{Almheiri:2019hni}. In this paper, we will often neglect this distinction and refer to $I$ itself as the island.} For this latter extremum, $S_\text{gen}$ is dominated by the area term in \eqref{Sgendef} and is therefore given approximately by the Bekenstein-Hawking  entropy of the black hole. The end result is the Page curve,
\be
S_\text{gen} = \text{min}( S_\text{rad}, S_\text{BH})\,,
\ee
and a Page time defined as the time when the two extrema trade places providing the smallest extremal value of $S_\text{gen}$.

In this paper we study black hole geometries in 1+1 dimensional dilaton gravity, which are asymptotically flat with an asymptotically linear dilaton field.  Linear dilaton spacetimes are familiar from constructions in string theory where they arise as holographic duals to some non-conformal theories. A prominent example is given by the near-horizon limit of NSNS fivebranes, which is a spacetime of the form
\be
{\bf R}^{1,5}\times {\bf R}_\phi \times S^3~,
\ee
where ${\bf R}_\phi$ denotes the direction along which the string theory dilaton is linear. This background is an $\alpha'$-exact solution of heterotic string theory \cite{Callan:1991at}. The dual field theory in this case is ${\cal N}=(1,1)$ supersymmetric Yang-Mills theory in six dimensions, which does not flow to a conventional QFT in the UV but rather to a non-local theory called little string theory \cite{Aharony:1998ub}. The dilaton gravity models studied in the present paper are in fact closely related to the above fivebrane background, as explained for example in \cite{Maldacena:1997cg}.  However, this will not play an important role in our discussion beyond exemplifying that linear dilaton spacetimes can serve as holographic backgrounds for a class of non-conformal theories. 

\begin{figure}[h!]
	\vspace{0.5cm}
	\begin{center}
		\begin{overpic}[width=.4\textwidth]{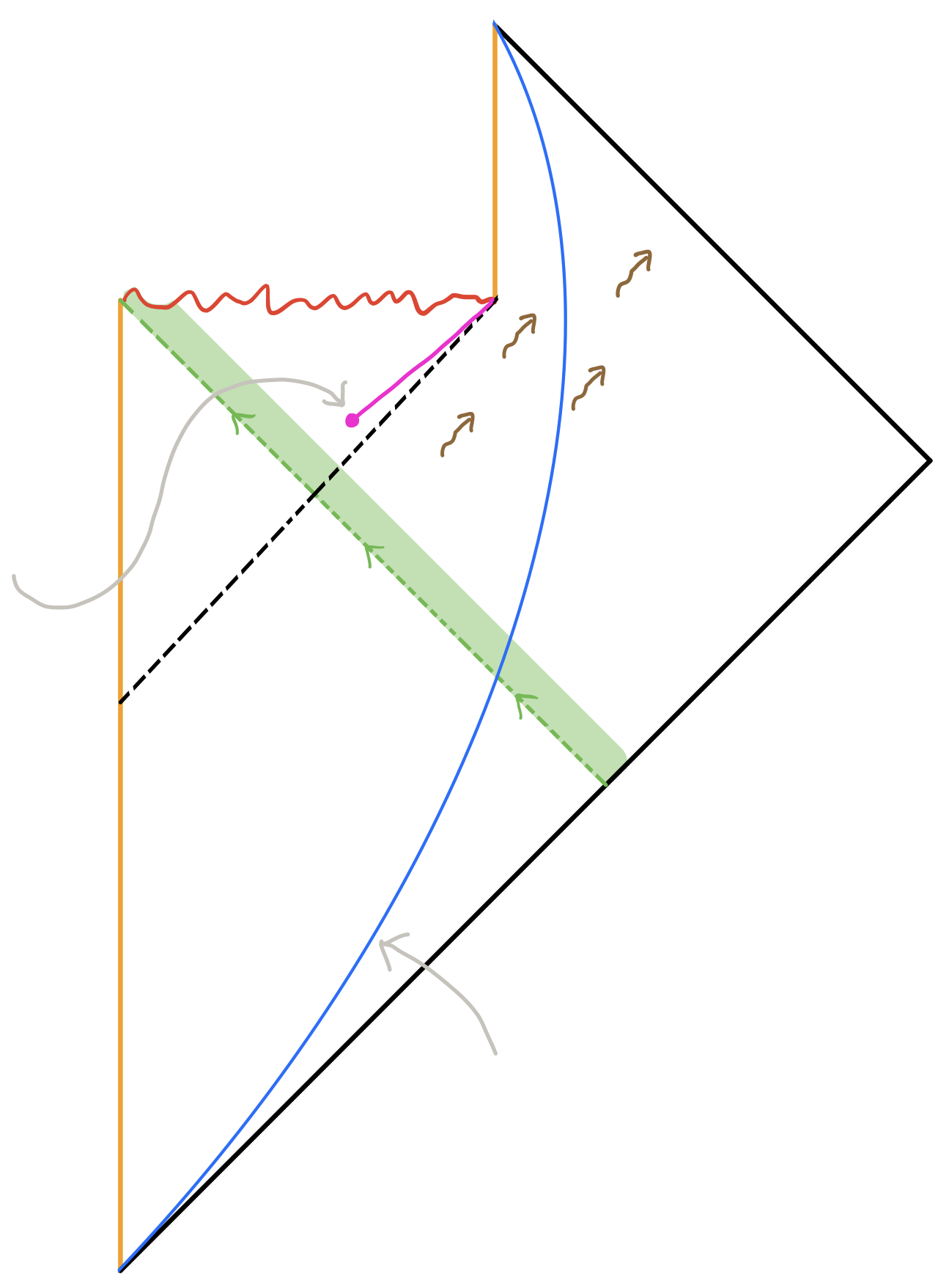}
			\put (41,15.5) {\footnotesize{Anchor}}
			\put (42.5,11.5) {\footnotesize{Curve}}
			\put (-3,58) {\footnotesize{Island}}
		\end{overpic}
	\end{center}
\label{CGHSrad} 
\caption{Penrose diagram of an evaporating RST black hole formed from collapsing matter (green). A timelike anchor curve separates the spacetime into interior and exterior regions. As time evolves along this curve, more and more Hawking radiation has passed through it on its way to future null infinity. The island moves with time along the purple curve inside the event horizon.
}
\end{figure}

The holographic dictionary for such linear dilaton backgrounds works in a similar way as in standard AdS/CFT, except the dual variables are defined in the asymptotic linear dilaton region instead of the AdS region in conventional holography (see \cite{Aharony:1998ub} for details).  In our computation we will place a timelike ``anchor curve'' at a fixed radial position far outside the black hole. 
In the gravitational theory the Hawking radiation emitted from the black hole will pass through the anchor curve as depicted in Figure~\ref{CGHSrad}. Hence, we do not need to artificially split our system into a QFT dual to the black hole plus an auxiliary system where the Hawking quanta are collected as in an AdS background. Instead the split is taken care of in a natural way by the anchor curve dividing the system into an ``inside'' part containing the black hole and and an ``outside'' region containing outgoing Hawking radiation. We will compute the entanglement entropy between the radiation that has passed through the anchor curve and all that remains inside, including the black hole itself, and see explicitly that it follows a Page curve as a function of time experienced by asymptotic observers who remain stationary with respect to the black hole.
The challenging aspect of the computation is the evaluation of the second term in~\eqref{Sgendef} for any given trial island $I$. To simplify this task, we follow~\cite{Almheiri:2019hni} and use AdS$_3$/CFT$_2$ duality to compute the von Neumann entropy of the bulk fields using a standard Ryu-Takayanagi prescription. We will come back to this in Sec.~\ref{Sgen}.

\section{The model}\label{model}
We start with the classical CGHS dilaton gravity action \cite{Callan:1992rs},
\be\label{CGHSaction}
I_\text{grav} = \f{1}{2\pi} \int \dd^2x\sqrt{-g}\,\e^{-2\phi}\left\{R+4(\nabla\phi)^2+4\lambda^2\right\}\,,
\ee
where $\phi$ is the dilaton field and $\lambda$ is a characteristic length scale that can be set to $\lambda=1$ by a rescaling of the two-dimensional coordinates. The vacuum solution is given by flat spacetime with a linear dilaton profile,
\be\label{lineardil}
\dd s^2 = -\dd \sigma^+ \dd \sigma^-\,,\quad \phi =\phi_0 -\sigma~,
\ee
where $\phi_0$ is an arbitrary constant that can be absorbed by a constant shift of the spatial coordinate 
$\sigma=\f12(\sigma^+{-}\sigma^-)$. 
The strength of the gravitational coupling is controlled by the dilaton field and becomes large as $\sigma$ tends to $-\infty$. 
The two-dimensional model can be viewed as a spherical reduction of a four-dimensional theory. In this case, the scale $\lambda$ is inherited from the parent theory and $\e^{-2\phi}$ is proportional to the area of the transverse 2-sphere in four-dimensional Planck units.

We will find it useful to employ so-called Kruskal coordinates, for which the metric in conformal gauge takes the form
\be\label{KruskalGauge}
\dd s^2 = - \e^{2\rho(x^+,x^-)} \dd x^+\dd x^-\,,
\ee
with the conformal factor equal to the dilaton $\rho=\phi$. In this coordinate system the equations of motion obtained from \eqref{CGHSaction} reduce to 
\be
\partial_+\partial_- \e^{-2\phi} +1= \partial_+^2  \e^{-2\phi}=\partial_-^2 \e^{-2\phi}=0~.
\ee
In the absence of matter fields, the dilaton gravity theory is non-dynamical and the general solution to the
above field equations, up to constant shifts of the $x^\pm$ coordinates, is given by
\be
\e^{-2\phi} = M- x^+x^-~,
\ee
where $M$ an integration constant. For $M=0$ we get back the vacuum solution \eqref{lineardil}, written in Kruskal coordinates. For $M<0$ the solution exhibits a naked singularity analogous to negative mass Schwarzschild solution in four dimensions. For $M>0$, a rescaling of the coordinates, $x^+ = \sqrt{M} v$ and $x^- = \sqrt{M} u$, gives the well-known two-dimensional `cigar' solution in Lorentzian signature~\cite{Mandal:1991tz,Witten:1991yr},
\be\label{WBH}
\dd s^2 = -\f{\dd v\,\dd u}{1-vu}~,
\ee
with a curvature singularity at $vu=1$ and a bifurcate event horizon at $vu=0$. The integration constant $M$ is proportional to the black hole mass,
\be
{\cal M} = \f{\lambda M}{\pi}~,
\ee
where we have temporarily restored the characteristic mass scale $\lambda$. 
The temperature of a CGHS black hole is independent of its mass,
\be\label{Htemp}
T = \f{\lambda}{2\pi}\>.
\ee
The Bekenstein-Hawking entropy, given by one quarter of the horizon area in Planck units in the original four dimensional theory, can be expressed in terms of the dilaton field evaluated at the horizon,
\be\label{eternalentropy}
S = 2\e^{-2\phi}\Big|_\text{Horizon}= 2M =  \f{2\pi\cal M}{\lambda}\>.
\ee
A purely two-dimensional argument leading to the dilaton dependence in \eqref{eternalentropy} is that while the area of the horizon is unity, the gravitational coupling constant is controlled by the dilaton as is apparent from \eqref{CGHSaction}, and this must be taken into account when evaluating the Bekenstein-Hawking entropy $S=\text{Area}/4G_N$.

\subsection{Coupling to matter}\label{matter}
In the original CGHS model \cite{Callan:1992rs}, the dilaton gravity sector is coupled to matter in the form of $N$ minimally coupled free scalars, with $N\gg 24$ so that semi-classical corrections are dominated by one-loop effects due to the matter fields. 
Here we will instead assume a strongly coupled matter sector described by a holographic two-dimensional CFT with large central charge~$c$
that has an AdS$_3$ gravitational dual. This is an important technical assumption which allows us to simply evaluate the von Neumann entropy of the CFT fields on a spacelike section but does not affect the gravitational sector. In particular, the theory still has solutions describing dynamical black holes formed from incoming matter energy-momentum.\footnote{When comparing to black holes in \cite{Callan:1992rs,Russo:1992ax} we make the identification $N=c$ and $\kappa =  c/12$.} 

Through the holographic dictionary, the two-dimensional central charge is related to three-dimensional gravitational quantities via the Brown-Henneaux formula,
\be
c = \f{3 L_3}{2G_{(3)}}~.
\ee
As discussed in Section~\ref{sec:dynamical} below, we can arrange our computation of the von Neumann entropy of the matter fields in such a way that we only have to deal with pure gravity in AdS$_3$ spacetime.

We are interested in semi-classical black holes with initial mass $M$ large compared to the scale set by the central charge of the matter CFT, for which there is a natural expansion parameter given by
\be
\epsilon\equiv\f{c}{48M}\ll 1~.
\label{epsilonparameter}
\ee
In most of what follows we work to leading non-trivial order in $\epsilon$, but to get started it is useful to consider the $\epsilon\rightarrow 0$ limit where semi-classical effects are turned off. In this limit, the gravitational field equation is sourced by the energy-momentum tensor of the two-dimensional matter CFT,\footnote{Our conventions match those of \cite{Polchinski:1998rq} except we are dealing with Lorentzian CFT. In particular, the classical energy-momentum tensor is defined as
\be
T^{\mu\nu} = -\f{4\pi}{\sqrt{-g}} \f{\delta S}{\delta g_{\mu\nu}}~,
\ee
and the normalization for the energy-momentum tensor therefore differs from the one used in \cite{Callan:1992rs} by a factor of 2.}
\be
4\e^{-2\phi}\Big[\nabla_\mu\nabla_\nu\phi - g_{\mu\nu}\big(\square \phi-(\nabla\phi)^2+1\big)\Big]=T_{\mu\nu}~,
\ee
while the equation of motion of the dilaton field is unaffected by the coupling to matter. The CFT energy momentum tensor has two non-trivial components $T_{++}(x^+)$ and $T_{--}(x^-)$, each of which only depends on one of the light cone coordinates.
The field equations take a particularly simple form in the Kruskal coordinates \eqref{KruskalGauge},
\be\label{CGHSnobackreactionEOM}
-\partial_+\partial_- \e^{-2\phi} =1 \,,\quad -2\partial_\pm^2  \e^{-2\phi}= T_{\pm\pm}~,
\ee
and the response to arbitrary incoming matter energy flux is easily obtained, 
\be\label{CGHSsolution}
\e^{-2\phi} = \e^{-2\rho}=F(x^+) - x^+\big[x^-+G(x^+)\big]~,
\ee
where
\be\label{FGCGHSeqs}
F'(x^+) = \f12{x^+ T_{++}(x^+)} ~,\qquad G'(x^+) = \f12{T_{++}(x^+)}~.
\ee
We take the energy-momentum tensor to have compact support in $x^+$ corresponding to a thin shell of infalling matter energy incident on the linear dilaton vacuum. For our purposes, the detailed form of the solution \eqref{CGHSsolution} is not needed, only the behaviour at early and late times, and we can therefore consider an idealised solution where two static configurations are patched together across an infinitely thin null shock wave,
\be
\label{kruskalmetric}
\e^{-2\phi(x^+,x^-)} = \e^{-2\rho(x^+,x^-)} =
\begin{cases} -x^+x^- & \text{if $x^{+}<x^+_0$}\,, \\ M-x^+\left( x^-+\f{M}{x^+_0}\right) & \text{if $x^{+}> x^+_0$}\,.
\end{cases}
\ee
A rescaling of the coordinates,
\be\label{rescaledcoords}
x^+ = x^+_0\, v~,\quad x^- = \f{M}{x^+_0}\,u~,
\ee
brings the metric and dilaton into the following simple form,
\be
\label{vumetric}
\e^{-2\rho(v,u)} = \f1M\,\e^{-2\phi(v,u)} =
\begin{cases} -v u & \text{if $v<1$}\,, \\ \left(1-v( u+1)\right) & \text{if $v>1$}\,.
\end{cases}
\ee
In the $v<1$ linear dilaton region, the change of coordinates,
\be
\label{omegas}
v=e^{\omega^+}\>,\qquad u=-e^{-\omega^-}\>,
\ee
brings the metric into manifestly flat form, $\rho(\omega^+,\omega^-)=0$, while a
set of coordinates, for which the metric is asymptotically Minkowskian in the $v>1$ region outside the 
shock wave, is given by 
\be
\label{sigmas}
v=e^{\sigma^+}\>,\qquad u=-1-e^{-\sigma^-}\>.
\ee
Time as measured by asymptotic observers at rest with respect to the black hole is
\be
\label{asymptotictime}
t=\f12 (\sigma^++\sigma^-)\>.
\ee

\subsection{Semi-classical black holes}\label{semiclassicalbackreac}

On a curved spacetime background, the energy-momentum tensor of the matter CFT is no longer traceless due to the conformal anomaly,
\be\label{traceanomaly}
\langle T_{\ \mu}^{\mu}\rangle = \f{c}{12} R~,
\ee
where $R$ is the Ricci scalar of the background metric. In two spacetime dimensions the continuity equation expressing energy-momentum conservation can be integrated using only \eqref{traceanomaly} as input \cite{Christensen:1977jc} leading to the following expressions in conformal coordinates,
\be\label{Tofrho}
\langle T_{+-} \rangle = -\f{c}{6} \partial_+\partial_-\rho~,\quad \langle T_{\pm\pm}\rangle=\f{c}{12}\left(2\partial_\pm^2 \rho-2(\partial_\pm\rho)^2 - t_\pm \right)~,
\ee 
where $t_\pm(x^\pm)$ are functions of integration determined by physical boundary conditions that reflect the matter quantum state.

The boundary functions $t_\pm$ are sensitive to the choice of coordinate system. This is to be expected since notions of positive frequency and normal ordering depend on the choice of time variable. Under a conformal reparametrization of the light-cone coordinates, $x^\pm\rightarrow y^\pm(x^\pm)$, the conformal factor of the metric transforms as 
\be
\rho(y^+,y^-) = \rho(x^+,x^-) -\f12 \log \f{\dd y^+}{\dd x^+} \f{\dd y^-}{\dd x^-}~.
\ee
When inserted in \eqref{Tofrho} this leads to the usual anomalous transformation of the energy-momentum tensor involving a Schwarzian derivative, 
\be\label{Ttrafo}
\left(\f{dy^\pm}{dx^\pm}\right)^2 T_{\pm\pm}(y^\pm) = T_{\pm\pm}(x^\pm) 
- \f{c}{12}\{y^\pm,x^\pm\}~,\quad \{y,x\} = \f{y'''}{y'}-\f32\f{(y'')^2}{(y')^2}~.
\ee
In order to preserve the form \eqref{Tofrho} for the energy-momentum tensor in the new coordinates, we effectively obtain a new function $t_\pm(y^\pm)$ related to the old one via
\be\label{littlettrafo}
\left(\f{dy^\pm}{dx^\pm}\right)^2 \, t_\pm(y^\pm) = t_\pm(x^\pm) + \{y^\pm,x^\pm\}~.
\ee
As an example, consider a black hole formed by gravitational collapse as in \eqref{vumetric}. At early advanced time before the arrival of the collapsing matter ({\it i.e.} $v<1$), we have a linear dilaton vacuum and vanishing energy-momentum tensor. The metric is manifestly flat in the $(\omega^+,\omega^-)$ coordinate system in \eqref{omegas} and it follows that $t_-(\omega^-)=0$. 
Upon transforming to the $(\sigma^+,\sigma^-)$ coordinate system \eqref{sigmas}, in which the metric is manifestly asymptotically flat, one finds non-vanishing outgoing energy flux at $\sigma^+\rightarrow \infty$,
\be\label{Hflux}
T_{--}(\sigma^-) = -\frac{c}{12} t_{-}(\sigma^-) = \f{c}{24}\left(1-\f{1}{(1+\e^{\sigma^-})^2}\right)~.
\ee
In \cite{Callan:1992rs} this expression was interpreted as the energy flux of Hawking radiation from the black hole as observed by an asymptotic observer. 
Energy conservation implies that a black hole emitting Hawking radiation loses mass. When the semi-classical expansion parameter $\epsilon$ in \eqref{epsilonparameter} has a small but finite value, the classical solution \eqref{CGHSsolution} is only valid on timescales that are short compared to the lifetime of the black hole, which is $t_\text{lifetime}= 1/\epsilon$ at leading order. The semi-classical back-reaction on the spacetime geometry due to Hawking emission matter can, however, be accounted for by adding to the classical action $I_\text{grav}$ in \eqref{CGHSaction} a non-local Polyakov term induced by matter quantum effects \cite{Callan:1992rs}, 
\be\label{IQdefinition}
I_Q =-\f{c}{12\pi} \int \dd x^+\dd x^- \partial_+\rho\partial_-\rho~,
\ee
written here in conformal coordinates.\footnote{While the non-local nature of $I_Q$ is not immediately apparent in the conformal gauge expression \eqref{IQdefinition}, it enters the formalism via the boundary functions $t_\pm$ in \eqref{Tofrho}.}
If we take $c\gg 24$ then $I_{Q}$ should be dominant compared to semi-classical contributions from the dilaton gravity sector. 
Further modifications to the theory are needed in order to find analytic solutions to the semi-classical equations of motion. We will follow the approach of \cite{Russo:1992ax} and add the following term,
\be\label{RST}
I_\text{RST} = \f{c}{48\pi }\int \dd^2 x \sqrt{-g} \phi R~,
\ee
to the semi-classical action, which is allowed by general covariance and does not disturb the classical ($\epsilon\rightarrow 0$) limit of the theory. The RST term $I_\text{RST}$ involves a factor of $c$ and therefore enters at the same order as the Polyakov term $I_{Q}$.

The resulting semi-classical field equations simplify dramatically when a new field variable is introduced,
\be\label{Omegaphirel}
\Omega = \e^{-2\phi} + \f{c}{24} \phi~,
\ee
but there are subtleties involved. In particular, the new field variable is bounded from below, $\Omega\ge \Omega_\text{crit}=\frac{c}{48} \left(1-\log \frac{c}{48}\right)$,
and when $\Omega\rightarrow \Omega_\text{crit}$ the gravitational coupling becomes strong in the semi-classical theory \cite{Russo:1992ax}. This has a suggestive physical interpretation, where $\Omega\rightarrow \Omega_\text{crit}$ represents a boundary of spacetime, analogous to the boundary at the origin of radial coordinates in the higher-dimensional theory from which the CGHS model is descended. 

One benefit of including the RST term \eqref{RST} is that semi-classical solutions of the full theory~$I_\text{grav}+I_Q+I_\text{RST}$ can be expressed in Kruskal coordinates \eqref{KruskalGauge}, where the field equations reduce to
\be\label{RSTEOM}
\partial_+\partial_- \Omega +1= 0\,,\quad - \partial_\pm^2  \Omega=\f{c}{24} t_\pm~,
\ee
with $t_\pm$ the same boundary functions as before.
The linear dilaton vacuum remains an exact solution of the semi-classical equations and takes the form
\be\label{RSTvacuum}
\Omega=-x^+ x^- - \f{c}{48}\log(-x^+x^-)~,
\ee
in the new field variable.
Notice that $t_\pm (x^\pm)\ne0$ even if this is the vacuum solution but this is because the metric is not manifestly flat in Kruskal coordinates. Transforming to a manifestly flat coordinate system $\sigma^\pm$ renders the functions $t_\pm(\sigma^\pm) =0$ as expected.

A two-sided eternal black hole solution is given by 
\be\label{Omegaeternal}
    \Omega = M(1-vu)
    + \Omega_\text{crit}
    \, ,
\ee
where we have rescaled the coordinates as in \eqref{WBH}. Here we find that in Kruskal coordinates that $t_\pm(x^\pm)=0$ but if we transform to coordinates for which the metric is manifestly asymptotically flat, 
\be\label{flatcoord}
v=e^{\sigma^+}\>,\qquad u=-e^{-\sigma^-}\>,
\ee
we find that $t_\pm(\sigma^\pm) = \f{1}{2}$. This corresponds to a flat space energy-momentum tensor $T_{\pm\pm}(\sigma^\pm) = \f{c}{24}$ which is exactly the energy-momentum tensor of a thermal gas of temperature $T=\f{1}{2\pi}$ which is the temperature of the eternal black hole. The outgoing energy flux carried by the Hawking radiation is matched by an incoming flux of thermal radiation at the same temperature as the Hawking temperature of the black hole.

Finally, consider the formation and subsequent evaporation of a dynamical black hole. As in the classical case without back-reaction, we imagine a situation where a short burst of matter energy is injected into a linear dilaton vacuum described by~\eqref{RSTvacuum}. The solution describing the full evolution of such a black hole can be found in~\cite{Russo:1992ax}. Here we are mainly interested in the geometry outside the collapsing matter shell, {\it i.e.} for $v>1$, where it takes the form
\be\label{dynamicalRST}
\dd s^2 = - M\e^{2\phi} \, \dd v \dd u~,\quad \Omega = M\big(1-v(u+1) - \epsilon \log(-Mvu)\big)~,
\ee
with $\phi$ and $\Omega$ related via \eqref{Omegaphirel}.

\section{Generalized entropy}\label{Sgen}

In order to derive a Page curve for these semi-classical black holes, we adapt the expression for the generalized entropy, 
\be\label{Sgendef2}
S_\text{gen} = \f{\text{Area}(I)}{4G_N} + S_\text{Bulk}[{\cal S}_{AI}] \,,
\ee
to the two-dimensional setting at hand. 
The first term on the right hand side involves the area of the transverse two-sphere evaluated locally at an island, and gives zero in the absence of an island. Comparing with the black hole entropy in \eqref{eternalentropy} yields $2e^{-2\phi(I)}$ as the area contribution of an island in the classical limit. The natural semi-classical extension of this expression, which gives zero in the absence of an island, is given by \be
\f{\text{Area}(I)}{4G_N} =2\big(\Omega(I)-\Omega_\text{crit}\big).
\ee

The second term on the right hand side in \eqref{Sgendef2} is universal and is the main focus of this section. It is the von Neumann entropy of the CFT matter fields on a spacelike surface ${\cal S}_{AI}$ that is bounded at one end by the island $I$ and at the other end by a point $A$ on a timelike anchor curve. We take the anchor curve to be a constant $\Omega$ curve with $\Omega=\Omega_A\gg M$ so that it is located well outside the black hole. For an eternal black hole \eqref{Omegaeternal} a curve of constant $\Omega$ is at a fixed spatial coordinate, $\sigma=\sigma_A$ in the manifestly asymptotically flat coordinate system \eqref{flatcoord}. For an evaporating black hole, the corresponding statement is no longer exact due to the $\log$ term in \eqref{dynamicalRST}. The spatial location of the anchor curve drifts in the asymptotic coordinates \eqref{sigmas} but for $\Omega_A\gg M$ the drift is extremely slow and can be ignored on time scales of order the black hole lifetime.
The final answer for $S_\text{Bulk}$ does not depend on which ${\cal S}_{AI}$ is chosen as long as it is a spacelike surface that connects $A$ and $I$.
In the absence of an island, the surface ${\cal S}_{AI}$ is instead bounded by $A$ at one end and a point on the boundary curve $\Omega=\Omega_\text{crit}$ at the other.

Following \cite{Almheiri:2019hni}, we compute the von Neumann entropy holographically by passing to a three-dimensional gravitational theory and evaluating the geodesic length between the points where $A$ and $I$ are embedded in the dual three-dimensional spacetime,
\be \label{eq:rt_3d}
S_\text{Bulk}[{\cal S}_{AI}] \simeq \f{\text{Length}}{4G_{(3)}}~.
\ee
The calculation is simplified if we arrange the embedding geometry to be pure AdS$_3$. This can be achieved in two steps.
The first step is to identify a set of light-cone coordinates 
\be
\dd s^2 = -\e^{2\rho} \dd y^+\dd y^-~.
\ee
where the integration functions $t_\pm(y^\pm)$ are zero.
The second step is to perform a Weyl rescaling of the two-dimensional metric that strips off the conformal factor $e^{2\rho}$.  
Then both $t_\pm(y^\pm)$ and the gravitational contribution the energy-momentum tensor in \eqref{Tofrho} vanish. In this case, the matter CFT is in a vacuum state and the dual three-dimensional geometry is empty AdS$_3$ spacetime. In Poincare coordinates the metric is 
\be
\dd s_3^2 = \f{L_3^2}{z^2}\,\big(\dd z^2 - \dd y^+\dd y^-\big)~,
\ee
and the geodesics are semi-circles centered on the holographic boundary. 
The Weyl transformation in step two above can be implemented as a coordinate transformation in three dimensions which maps the regulated holographic boundary to a surface,  
\be
z=\delta \, \e^{-\rho(y^+,y^-)}~,
\ee
that depends on dynamical input from the two-dimensional matter theory. Here $\delta$ is a UV cutoff parameter. 

A standard calculation involving AdS$_3$ geodesics then leads to the following result for the holographic entropy,
\be\label{Bulkresult}
S_\text{Bulk}[{\cal S}_{AI}] = \f{c}{6}\log \left[ d(A,I)^2 \e^{\rho(A)}\e^{\rho(I)}\right]\Big|_{t_\pm =0}~,
\ee
where $d(A,I)$ is the two-dimensional distance measured between the points $A$ and $I$ in the flat metric 
$\dd s^2_\text{flat} = -\dd y^+\dd y^-$ and the subscript is a reminder that the formula should be evaluated in coordinates 
for which $t_\pm(y^\pm)=0$. We have dropped the UV cutoff from the formula as it just contributes an additive constant. 


\section{Page curves}\label{sec:pages}

We now have everything in place to calculate generalized entropy in the RST model using the QRT prescription.
Our primary goal is to obtain the Page curve of an evaporating black hole that has a finite lifetime but first we 
carry out the corresponding calculation for a semi-classical eternal black hole. This provides a first test involving a black hole
in asymptotically flat spacetime which turns out to be considerably simpler than the evaporating case.

\subsection{Eternal black hole}\label{sec:eternal}

A semi-classical eternal black hole in asymptotically flat spacetime is supported by a thermal gas of incoming radiation that maintains the mass of the black hole against the energy loss to Hawking radiation. The two-dimensional black holes studied in this paper all have  temperature $T=\f{1}{2\pi}$ and the thermal gas must be at the same temperature. As was noted below \eqref{Omegaeternal}, the energy flux outside an eternal semi-classical RST black hole is 
\be
T_{\pm\pm}(\sigma^\pm) = \f{c}{24}~,
\ee
when evaluated in manifestly asymptotically flat coordinates and
this is precisely the energy-momentum tensor of a thermal CFT at a temperature of $T=\f{1}{2\pi}$.
It was also noted that $t_+(v)=t_-(u)=0$ for an eternal black hole and therefore $(v,u)$ is the appropriate set of coordinates to use when evaluating $S_\text{Bulk}$ in~\eqref{Bulkresult}. The relevant matter CFT vacuum state is the Hartle-Hawking state where positive frequency modes are determined with respect to time in Kruskal coordinates rather than asymptotic Minkowski time. 

\begin{figure}
\centering
\includegraphics[width=0.7\textwidth]{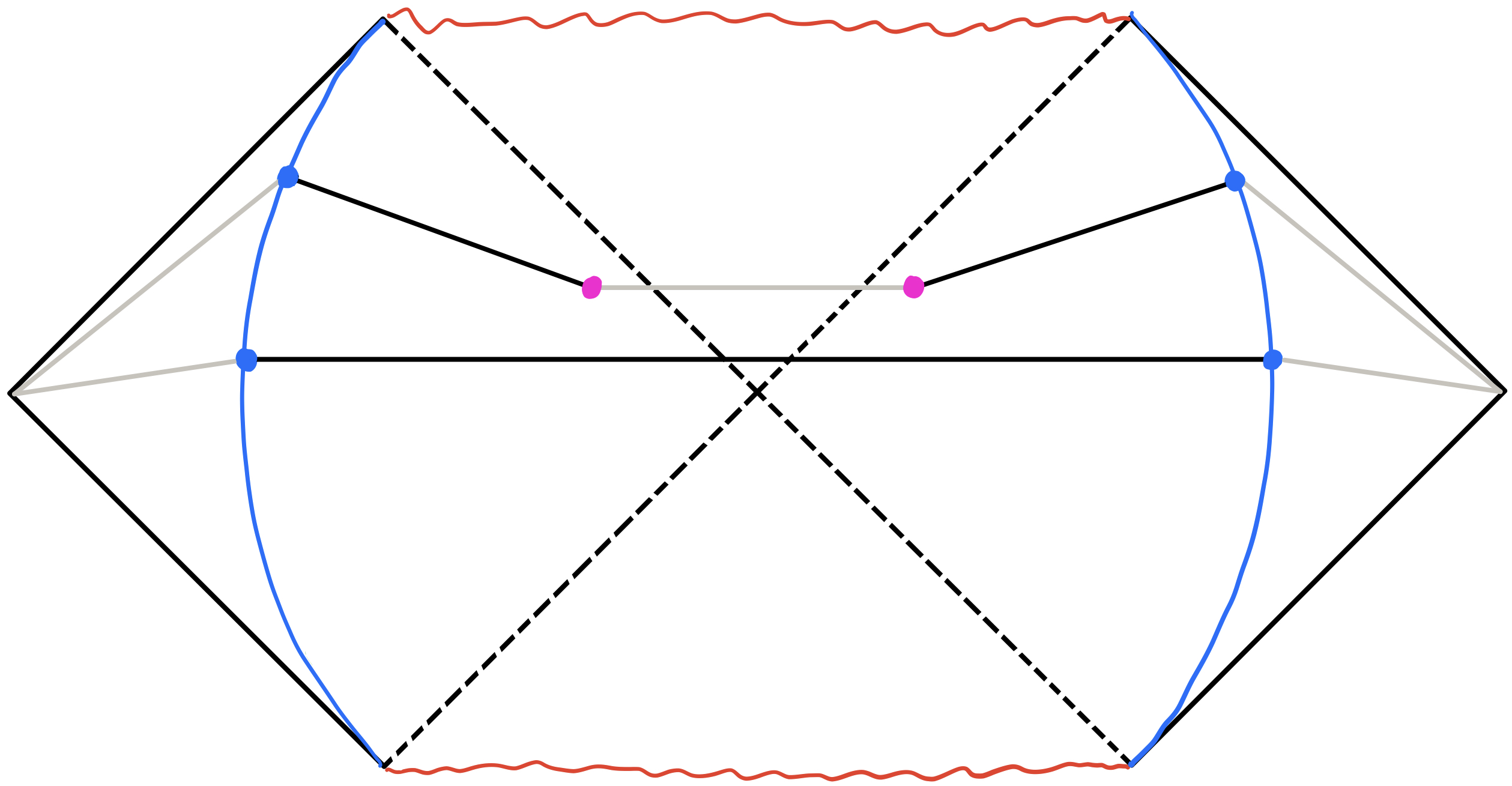}
\caption{\label{eternal} A Penrose diagram of an eternal black hole. A pair of timelike anchor curves (blue curves) separates the spacetime into an interior and two exteriors. The two spatial hypersurfaces intersect the anchor curves at different times. On the late time surface the generalized entropy is dominated by the area term associated to the islands denoted by purple dots.}
\end{figure}

The eternal black hole is two sided and we place a timelike anchor curve in each asymptotic region. For simplicity, we assume that our anchor points lie symmetrically on the anchor curves, as shown in Figure~\ref{eternal}. Then each anchor point has a mirror anchor point  (denoted by superscript $m$) in the other exterior region, which is related to the original point by $(v,u)^m = (u,v)$. 
We also take the black hole mass to be large compared to the scale set by the matter central charge, so that $\epsilon=\f{c}{48M}\ll 1$, and the anchor curves to be located in the linear dilaton region, so that $\Omega_A\gg M$. With these assumptions in place the semi-classical field variables are well approximated by their classical counterparts in all regions of interest and our calculations simplify.

Inspired by \cite{Penington:2019npb,Almheiri:2019hni}, we now perform two calculations: One with no islands, and one with a single island on each side. Consider first the no-island scenario. In this case, the area term of the generalized entropy is by definition zero, as $I$ is empty. The von Neumann entropy of the bulk fields is non-vanishing and given by the length of the geodesic in AdS$_3$ that connects the two mirrored anchor points. This means that we can directly apply~\eqref{Bulkresult} but with $I$ replaced by $A^m$, as indicated in Figure~\ref{eternal},
\be
S_\text{bulk} = \f{c}{12}\log\big[(v_A-v_{A^m})^2(u_A-u_{A^m})^2 \, e^{2\rho(v_A,u_A)}e^{2\rho(v_{A^m},u_{A^m})}\big] \>,
\ee
where $(v_A,u_A)$ denotes an anchor point on the curve on the right in the figure. 
The anchor curves are assumed to be located well outside the black hole where the conformal factor is well approximated by its classical value, 
\be\label{classapprox}
e^{2\rho(v,u)} \approx \f{1}{1-vu} \>.
\ee
The bulk entropy then takes a simple form,
\be\label{Sgrowth}
S_\text{bulk} = \f{c}{12}\log\f{(v_A-u_A)^4}{(1-v_Au_A)^2} \approx \f{c}{3}t_A  \>,
\ee
where $t_A$ is asymptotic time, measured by an observer on the anchor curve, and the asymptotically flat coordinates $(t,\sigma)$ are related to the $(v,u)$ coordinates via,
\be
v =e^{t+\sigma}\>,\qquad u =-e^{-t+\sigma}\>.
\ee
Corrections to this result are either exponentially suppressed (by factors of $e^{-2t_A}$ or $e^{-2\sigma_A}$) or subleading in powers of $\epsilon$, or both. Our computation includes, by construction, the entropy of the radiation emitted on both sides of the black hole and
we note that the entropy growth rate in \eqref{Sgrowth} is precisely twice the rate that was obtained in \cite{Fiola:1994ir} for the entanglement entropy of radiation emitted to one side. 

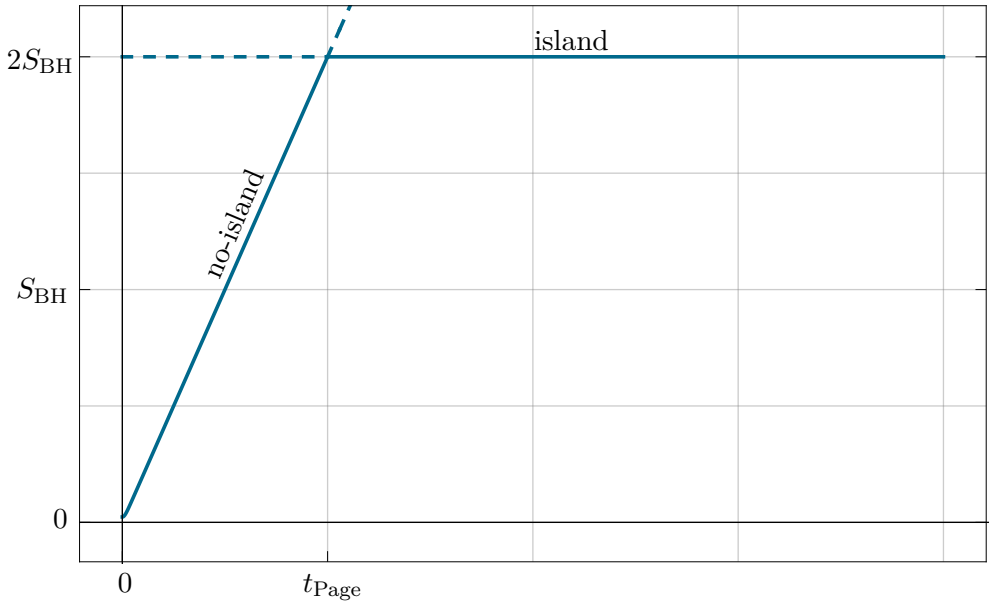
\begin{figure}[h!]
    \centering
    \begin{overpic}[]{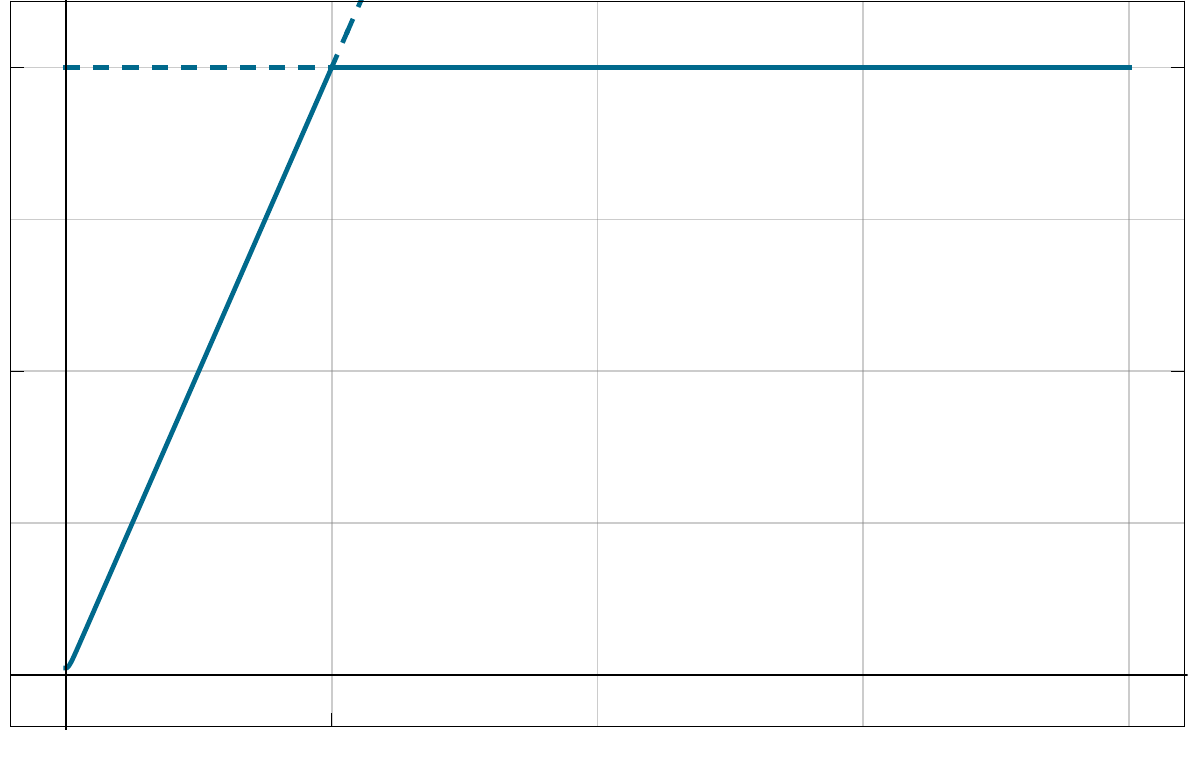}
        \put (5,1) {\small{$0$}}
        \put (25,1) {\small{$t_\text{Page}$}}
        \put (-2,8) {\small{$0$}}
        \put (-6,33) {\small{$S_\text{BH}$}}
        \put (-7,58) {\small{$2 S_\text{BH}$}}
        \put (14,35) {\rotatebox{67}{\small{no-island}}}
        \put (50,60) {\small{island}}
    \end{overpic}
    \caption{\label{eternalpagecurve} Page curve for the eternal RST black hole with $t_{\text{Page}}=6S_{\text{BH}}/c$. The graph plots $S_\text{gen}-\f{c}{3}\sigma_A$ as a function of retarded time on the anchor curve.}
\end{figure}

We now repeat the calculation with symmetrically placed islands at $I=( v_I,  u_I)$ and $I^m = (u_I,v_I)$, as indicated in Figure~\ref{eternal}. In this case, the area term in the generalized entropy \eqref{Sgendef2} is non-vanishing and bulk term involves geodesics in AdS$_3$ that connect the anchor point and corresponding island on each side of the black hole. The contributions from the two sides of the black hole are identical and add up to 
\be\label{Sislands}
S_\text{gen}^\text{island} = 4M(1- v_I  u_I) +\f{c}{6}\log\f{(v_A- v_I)^2(u_A-u_I)^2}{(1-v_Au_A)(1- v_I  u_I)}~,
\ee
where we have used \eqref{Omegaeternal} for the semi-classical area function $\Omega(I)-\Omega_\text{crit}$. Both the anchor point and the island are assumed to lie in a region where the classical approximation \eqref{classapprox} can be used for the conformal factor. This is automatically satisfied for an anchor point outside a large mass black hole and we will check ex post facto that it also holds for the island.
Extremizing over $(v_I,u_I)$ and working to leading order in $\epsilon\ll 1$, we obtain
\be
\label{vIuIeternal}
\f{u_I}{v_I} = \f{u_A}{v_A}~,\quad v_I \approx -\f{4\epsilon}{u_A}~.
\ee
This is a saddle point and not a minimum. However, the QRT prescription instructs us find all extrema and select the one that gives the lowest value for the generalized entropy. 
Inserting the leading order saddle point values for $v_I$ and $u_I$ into \eqref{Sislands} gives 
\be\label{eternalafterpage}
S_\text{gen}^\text{island}=4M + \f{c}3 \,\sigma_A+\ldots~.
\ee
Comparing to the no-island result in \eqref{Sgrowth} shows that for $t_A>\sigma_A + \f1{4\epsilon}$ the generalized entropy is dominated by the island configuration. The $\sigma_A$ term accounts for the time it takes for the Hawking radiation to travel from the black hole to the anchor curve. Correcting for this, we obtain
\be\label{eternalpagetime}
t_\text{Page} = \frac1{4\epsilon}=\f{12M}{c} \>,
\ee
for the Page time of an eternal RST black hole. The Page curve is drawn in Figure~\ref{eternalpagecurve}.

For a two-sided black hole in AdS$_2$ gravity, the island and its mirror were found to be outside the event horizon \cite{Almheiri:2019yqk}. This remains true here as well. The island saddle point \eqref{vIuIeternal} is outside the event horizon but inside the stretched horizon, with the proper distance between island and event horizon given by a tiny number,
\be
d_I\approx 4\epsilon\,\sqrt{\f{M}{\Omega_A}}\>.
\ee
The fact that the island is close to the event horizon justifies using the classical approximation for the conformal factor in \eqref{Sislands}, as promised. 
For another perspective on the location of the island, consider an observer sitting on the anchor curve who sends an ingoing light signal to the island. A straightforward calculation shows that in order to be received at an island at $(v_I,u_I)$, that corresponds to an anchor point at time $t_A$, the signal must be emitted from the anchor curve at an earlier time $t_A^\text{obs}$, such that
\be
t_A -t_A^\text{obs}=2\sigma_A + t_s~,
\ee
where $t_s = \log{(\f1{4\epsilon})}$ is the scrambling time. This is in line with a similar result in \cite{Almheiri:2019yqk}. However, because our black hole is in asymptotically flat spacetime and not AdS$_2$ the time difference $t_A -t_A^\text{obs}$ explicitly depends on the location of the anchor curve.

\subsection{Dynamical black hole}\label{sec:dynamical}
We now turn our attention to dynamical black holes and compute a Page curve for a black hole that is formed by gravitational collapse of matter and then gradually evaporates due to Hawking emission. The steps in the calculation are the same as before, {\it i.e.} to find extrema of the generalized entropy with and without an island and determine which one gives the minimum value. The area term in the generalized entropy can be read off directly from the semi-classical black hole solution but the remaining bulk term requires more work.

In the holographic evaluation of the bulk entropy term in \eqref{Bulkresult} we are instructed to identify light-cone coordinates where the $t_\pm$ contribution to the two-dimensional matter energy momentum tensor is zero. The correct choice is the $(\omega^+,\omega^-)$ system in \eqref{omegas} where the metric is manifestly flat in the initial linear dilaton region before the matter shell collapses to form the black hole. These coordinates are suitable for the evaluation of \eqref{Bulkresult} when calculating the generalized entropy on a trial surface in the linear dilaton vacuum where the CFT is manifestly in its vacuum state and the three-dimensional holographic dual is pure AdS$_3$. Of course, a dynamical black hole is not the linear dilaton vacuum and $t_+(\omega^+)$ is non-vanishing due to the incoming energy flux that forms the black hole. There is, however, a simple way around this problem. Following \cite{Fiola:1994ir}, we take the incoming matter to be described by a coherent state built on the vacuum state of inertial observers at past null infinity. As shown in~\cite{Fiola:1994ir}, the von Neumann entropy of such a state is identical to the von Neumann entropy of the vacuum state. As a result, we can use the AdS$_3$/CFT$_2$ Ryu-Takayanagi prescription \eqref{Bulkresult} to calculate the bulk term in the generalized entropy, provided we use the coordinate system that corresponds to the CFT in its vacuum state. This means in particular, that we are instructed to calculate the two-dimensional distance $d(A,I)$ in $(\omega^+,\omega^-)$ coordinates. 

\begin{figure}[t]
\centering
\includegraphics[width=0.4\textwidth]{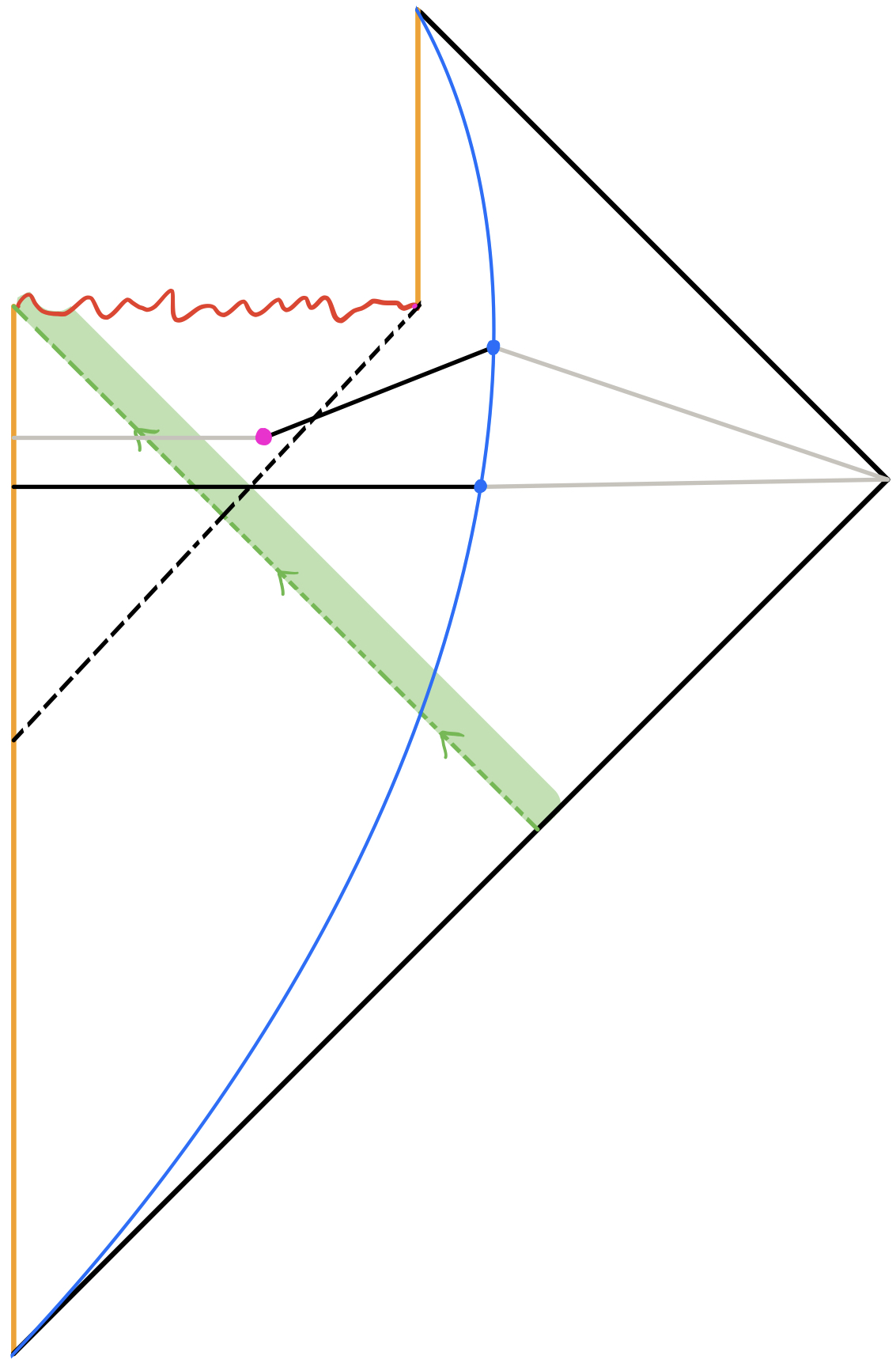}
\caption{\label{island_rst} Penrose diagram of a dynamical RST black hole with two spacelike hypersurfaces indicated, one before the Page time and the other after, corresponding to the no-island and island configurations, respectively.}
\end{figure}

The generalized entropy is to be computed for the two competing configurations, with and without an island, indicated in the Penrose diagram in Figure~\ref{island_rst}. The final result is the one that gives a smaller value for the entropy.

\subsubsection{Island configuration}
Let us start by determining the generalized entropy for an island configuration,
\begin{align}\label{Sgendyn}
S_\text{gen}^\text{island} = &\ 2M\big(1-v_I(1+u_I)-\epsilon\log{(-Mv_Iu_I)}\big)\\
 & {+}\f{c}{12} \log\left[\Big(\log\f{v_A}{v_I}\log\f{u_A}{u_I}\Big)^2\f{ v_Au_A}{(1-v_A(1+u_A))}\f{v_I u_I}{(1-v_I(1+u_I))} \right] ,\nonumber
\label{Sgendyn}
\end{align}
where we have used \eqref{dynamicalRST} for the area term $2(\Omega(I)-\Omega_\text{crit})$ and the coordinate distance $d(A,I)=\sqrt{-\Delta\omega^+\Delta\omega^-}$ has been expressed in $(v,u)$ coordinates. We are assuming that the island is located outside the infalling shell of matter and that both the anchor point and the island lie in a region where a classical approximation can be used for the conformal factor of the dynamical black hole metric. The anchor point is by assumption far outside the black hole where the classical approximation is always valid. It turns out to also be valid for the island for much of the lifetime of an evaporating black hole provided it starts out with a large enough mass but it will fail towards the end of the lifetime when the black hole has evaporated down to a small size. 

Extremizing \eqref{Sgendyn} over $(v_I,u_I)$ yields the following two conditions,
\begin{eqnarray}
0&=& -2M(1+u_I)+\f{c}{12v_I}\,\f{1+u_I}{(1-v_I(1+u_I))} + \f{c}{24v_I} - \f{c}{6v_I\log\big(\f{v_A}{v_I}\big)}\,, \\
0&=& -2Mv_I+\f{c}{12u_I}\,\f{v_I}{(1-v_I(1+u_I))} + \f{c}{24u_I} - \f{c}{6u_I\log\big(\f{u_A}{u_I}\big)}\,.
\end{eqnarray}
In order to solve for the location of the island we make the simplifying assumption $\log(\f{v_A}{v_I})\gg 1$, which allows us to drop the last term on the right in the top equation, and later on we verify the self-consistency of this assumption. The resulting equations can be rearranged as 
\be\label{simpleeqs}
(v_I(1+u_I))^2-(1-\epsilon)\,v_I(1+u_I)+\epsilon=0 \>\quad\text{and}\quad \log\big(\f{u_A}{u_I}\big)=4(1+u_I) \>.
\ee
One of the two solutions of the quadratic equation for $v_I(1+u_I)$ corresponds to an island in the near horizon region,
\be
u_I=-1+\frac{\epsilon}{v_I}+O(\epsilon^2)\>.
\ee
The other solution has the island near the black hole singularity and is unphysical. 
Inserting the island solution into the remaining equation in \eqref{simpleeqs} we find
\be
u_A=-1-\f{3\epsilon}{v_I}+O(\epsilon^2)\>.
\ee
In terms of the asymptotic coordinates \eqref{sigmas} we have
\be 
u_A= -1 -e^{\sigma_A-t_A}\>, \qquad v_A= e^{\sigma_A+t_A} \>.
\ee
The above relations imply that
\be\label{vAvIrelation}
\log\big(\f{v_A}{v_I}\big)\approx 2\sigma_A+\log\big(\f{1}{\epsilon}\big)\gg 1 \>,
\ee
so the simplifying assumption that we used to obtain the island solution is indeed justified.
We also assumed in the calculation that the island is located at $v_I>1$ and 
this turns out to be valid when $t_A-\sigma_A\gtrsim \log(1/\epsilon)$. 
The expression for the outgoing energy flux \eqref{Hflux} reveals that the first Hawking radiation passes through the anchor curve at $t_A-\sigma_A\approx 0$ and the island solution is already valid within a time of order the scrambling time after that. 

We can again probe the location of the island by considering an observer sitting on the anchor curve who sends an ingoing light signal. The relation between $v_I$ and $v_A$ in \eqref{vAvIrelation} implies that in order to be received at an island at $(v_I,u_I)$, the signal must be emitted from the anchor curve at a time $t_A^\text{obs}$, such that $t_A -t_A^\text{obs}=2\sigma_A + t_s$ with $t_s$ the black hole scrambling time. Earlier, we found the same result for an island just outside the event horizon of an eternal black hole. Here the island is inside the black hole but still located very close to the event horizon. 

Finally, inserting the leading order saddle point values for $v_I$ and $u_I$ into \eqref{Sgendyn} gives
\be\label{islandresult}
S_\text{gen}^\text{island} = 2M - \f{c}{24}(t_A-\sigma_A)+\ldots \>,
\ee
as a function of the retarded time at the anchor curve. Although the time-dependent contribution is initially of order $\epsilon$, compared to the leading order area term, it is important to keep in mind that this contribution grows with time, and eventually becomes comparable to the leading order result. This expression for the generalized entropy, which is valid when a time of order the scrambling time has passed after the first Hawking radiation emerges into the outside region beyond the anchor curve, is to be compared to the contribution from a no-island configuration that we now turn our attention to. 

\subsubsection{No-island configuration}

In the absence of an island, the spacelike surface ${\cal S}_{AI}$ extends from the anchor curve to the semi-classical boundary at $\Omega=\Omega_\text{crit}$. The gravitational coupling becomes strong at the semi-classical boundary and it is not a priori clear how to proceed. The validity of the semi-classical black hole solution indeed breaks down near the boundary but from a higher dimensional perspective this has a simple interpretation in terms of the area of the transverse two-sphere going to zero. We are primarily interested in the dependence of the bulk entropy \eqref{Bulkresult} on asymptotic time and this will not be greatly affected by the detailed conditions imposed at the origin. 
This can be seen by adopting a simple prescription for the strong coupling region and then checking that a change in the prescription does not change the leading order result at late times on the anchor curve. 

In the following we let the spacelike surface ${\cal S}_{AI}$ end at a fixed reference point $(v_0,u_0)$ on the boundary curve for all anchor points. In other words, we will simply ignore any adjustment of the endpoint at the semi-classical boundary in response to changing the anchor point. We take the reference point to be in the $v<1$ linear dilaton region inside the matter shock wave, {\it i.e.} with $u_0=-\epsilon/v_0$ for some $v_0<1$. The generalized entropy is given by
\begin{eqnarray}
S_\text{gen}^\text{no-island} &=& \f{c}{12}\log\left[\Big(\log\f{v_A}{v_0}\log\f{u_A}{u_0}\Big)^2\f{ v_Au_A}{(1-v_A(1+u_A))}\right] \nonumber \\
&=& \f{c}{12}(t_A-\sigma_A)+\ldots  \>,
\label{noislandresult}
\end{eqnarray}
up to logarithmic correction terms. In particular, all dependence on $v_0$ is contained in the sub-leading terms that we have dropped.
This expression is valid for retarded time of order the scrambling time and onwards.

Comparing the island and no-island results in \eqref{islandresult} and \eqref{noislandresult}, respectively, shows that for retarded time 
$t_A-\sigma_A > \f1{3\epsilon}$ the generalized entropy will be dominated by the island configuration. The Page time for a dynamical RST black hole is one third of the black hole lifetime,
\be\label{RSTpagetime}
t_\text{Page} = \frac1{3\epsilon}=\f{16M}{c} \>.
\ee
The corresponding Page curve is drawn in Figure~\ref{backreactpage}.

\section{Discussion}\label{sec:conclusion}
By assuming a QRT formula we have explicitly obtained Page curves for semi-classical black holes in asymptotically flat spacetime in a two-dimensional dilaton gravity model. This includes both an eternal black hole, supported by an incoming energy flux matching the outgoing Hawking flux, and a black hole formed by gravitational collapse that gradually evaporates. In both cases, the generalized entropy is minimised at early times by the bulk von Neumann entropy of the two-dimensional matter CFT but at a Page time the system crosses over to a configuration where the minimum generalized entropy includes a non-trivial Ryu-Takayanagi area term associated with an island near the black hole horizon. This is consistent with earlier results obtained for AdS$_2$ black holes but our computation offers a rather clean physical picture and offers evidence that the QRT prescription applies beyond asymptotically AdS spacetimes.

For the eternal black hole we confirm the appearance of an island outside the horizon, whereas for an evaporating black hole the island is always inside the horizon. Towards the end of evaporation the island appears to melt together with both singularity and horizon at the black hole endpoint. Our semi-classical calculation is, however, only reliable as long as the black hole mass remains large compared to the scale set by the central charge of the matter CFT. 

For the evaporating black hole the Page time comes out to be 1/3 of the black hole lifetime. This fits with the following simple reasoning \cite{Zurek:1982zz}. Radiating into a cold surrounding space is an irreversible process and the entropy of the radiation grows at twice the evaporation rate of the black hole. This can be seen more explicitly by considering the relation between internal energy and entropy of the gas emitted by a black body in two dimensions, 
\be
S =  \f{2 U}{T}~.
\ee
We can compute the ratio of the entropy of the gas that the black hole emits compared to the entropy lost by the black hole. The black hole satisfies the first law $\Delta S_\text{BH} = -\Delta M/T$ where $\Delta M$ is the mass lost by the black hole in a given time interval. The entropy increase of the gas radiated is $\Delta S_\text{gas} = 2 \Delta U / T$ where $\Delta U$ is the energy of the emitted gas, which must equal $\Delta M$, and we get $\Delta S_\text{gas} = -2 \Delta S_\text{BH}$.

Our results rely on two dimensional conformal methods. On a technical level this is reflected in the fact that we can conveniently obtain the bulk contribution to the entropy via a AdS$_{3}/$CFT$_{2}$ computation. It is furthermore convenient that in two dimensions there are no grey body factors. On a more fundamental level, we are able to account for radiation and back-reaction in the RST model by working in a conformal gauge and explicitly using the conformal anomaly in two dimensions. Although there is no known analogue of the RST model in higher dimensions, our result suggests that a QRT like prescription may also work for other non-AdS black holes, especially ones that can be embedded into a boundary theory with a higher-dimensional AdS dual.

\subsection*{Acknowledgments}
We thank Valentina Giangreco M. Puletti for stimulating discussions.
LT would like to thank the Kavli Institute for Theoretical Physics at UC Santa Barbara for hospitality during the completion of this work.  This research was supported in part by the Icelandic Research Fund under grants 185371-051 and 195970-051, by the University of Iceland Research Fund, and by the National Science Foundation under grant NSF PHY-1748958. FFG is a Postdoctoral Fellow of the Research Foundation - Flanders (FWO). FFG is also supported by the KU Leuven C1 grant ZKD1118 C16/16/005.

\bibliographystyle{JHEP}
\bibliography{ref_comp}
\end{document}